\documentclass[%
 reprint,
 amsmath,amssymb,aps,prl
]{revtex4-2}

\usepackage{graphicx}
\usepackage{dcolumn}
\usepackage{bm}
\usepackage{amsmath}
\usepackage{hyperref}

\begin{document}

\preprint{APS/123-QED}

\title{Stable eigenmodes of the Large-scale circulation in Turbulent Convection}

\author{Yin Wang$^1$}
 \email{ywang3@pppl.gov}
\author{Pik-Yin Lai$^2$}
\author{Penger Tong$^3$}
\affiliation{$^1$Princeton Plasma Physics Laboratory, Princeton University, Princeton, New Jersey 08543, USA \\
             $^2$Department of Physics and Center for Complex Systems, National Central University, Chungli, Taiwan 320, R.O.C. \\
             $^3$Department of Physics, Hong Kong University of Science and Technology, Hong Kong}%



\date{\today}

\begin{abstract}
Turbulent Rayleigh-Bénard convection (RBC) is a canonical system for studying buoyancy-driven turbulence. Its defining feature is a system-size large-scale circulation (LSC), whose dynamics reflect the coupling between coherent motion and turbulent fluctuations. Starting from the Boussinesq Navier-Stokes and temperature equations, we derive three coupled stochastic equations for the LSC angular velocity and its associated dipolar temperature field.
The deterministic dynamics contain an unstable no-LSC set and two symmetry-related stable finite-LSC states.
Linearization about either finite-LSC state yields one purely relaxational eigenmode and a pair of damped oscillatory eigenmodes whose frequency is set by the LSC turnover rate.
Turbulent fluctuations continually excite the latter, providing a dynamical origin for the observed quasi-periodic oscillations.
Experiments in a thin-disk cell directly resolve both eigenmodes and show their predicted Rayleigh-number scalings.
The same stochastic dynamics also provide a route to rare cessations and reversals.
This theory–experiment framework provides a tractable example of how coherent circulation states, intrinsic modes, and fluctuation-driven transitions emerge in turbulent convection, with broader relevance to geophysical and astrophysical flows.
\end{abstract}

\maketitle

Turbulent convective flows can spontaneously organize into coherent motions on scales far larger than the fluctuations from which they emerge.
Understanding how such large-scale structures are maintained, what determines their stability and intrinsic dynamical modes, and how they respond to turbulent fluctuations is a fundamental problem in nonequilibrium fluid dynamics.
Related questions arise in geophysical and astrophysical flows,
where coherent circulation states coexist with strong fluctuations and may possess multiple states or long-lived oscillatory modes.
For example, the Atlantic eridional overturning interacts with the density field it transport and can exhibit multiple circulation states and abrupt transitions~\cite{stommel1961,weijer2019}.
In the solar convection zone, global inertial modes with well-defined frequencies can be stochastically excited by turbulent convection~\cite{loptien2018,bekki2022}.
These examples raise a broader question: how do coherent circulation states and their intrinsic dynamical modes emerge, remain observable, and undergo transitions in strongly turbulent environments?

Turbulent Rayleigh--B\'enard convection (RBC) provides a canonical setting for addressing these questions quantitatively~\cite{ahlers2009}.
At sufficiently large Rayleigh number $Ra$, the flow spontaneously organizes into a system-spanning large-scale circulation (LSC) that exhibits coherent quasi-periodic oscillations, intermittent cessations, and reversals~\cite{niemela2001,funfschilling2004,sun2005,brown2005,araujo2005,brown2009,podvin2012,wang2018}.
In cylindrical cells, the oscillations have been attributed to three-dimensional (3D) torsional and sloshing motions of the LSC and described phenomenologically as a damped oscillator produced by geometry-dependent restoring forces~\cite{funfschilling2004,brown2007,xi2009,brown2009}.
Other models instead attribute the oscillation to intrinsic bulk-flow precession~\cite{resagk2006}, delayed coupling of thermal-plume emission~\cite{song2011a}, or travelling-wave-like velocity structures with frequencies set by the mean circulation rate~\cite{li2026}.
What remains missing is a dynamical framework that derives the underlying LSC states and their intrinsic stability modes from the coupled momentum and temperature dynamics and enables quantitative tests of both their decay rates and frequencies.

Here we address this question in a thin-disk convection cell designed to isolate the essential LSC dynamics.
Its quasi-2D confinement suppresses torsional and sloshing modes, while the circular cross section eliminates corner flows, yet the characteristic coherent oscillations, cessations, and reversals persist.
Starting from the Boussinesq Navier–Stokes and temperature equations, we derive three coupled stochastic equations for the LSC angular velocity and the amplitude and orientation of its dipolar temperature field.
The deterministic dynamics contain an unstable no-LSC set and two symmetry-related stable finite-LSC states. Linearization about either finite-LSC state yields one purely relaxational eigenmode and a pair of damped oscillatory eigenmodes.
Excitations of the oscillatory eigenmodes by turbulent fluctuations produce the observed quasi-periodic oscillation, whose frequency is set by the LSC turnover rate.
Unlike phenomenological models, the present framework predicts both the fixed-point structure and the real and imaginary parts of the stability eigenvalues from the governing dynamics.
These predictions, including their $Ra$ scalings, are tested directly against experiments, while cessations and reversals arise as fluctuation-driven transitions within the same framework.

\begin{figure}
\centerline{\includegraphics[width=0.33\textwidth]{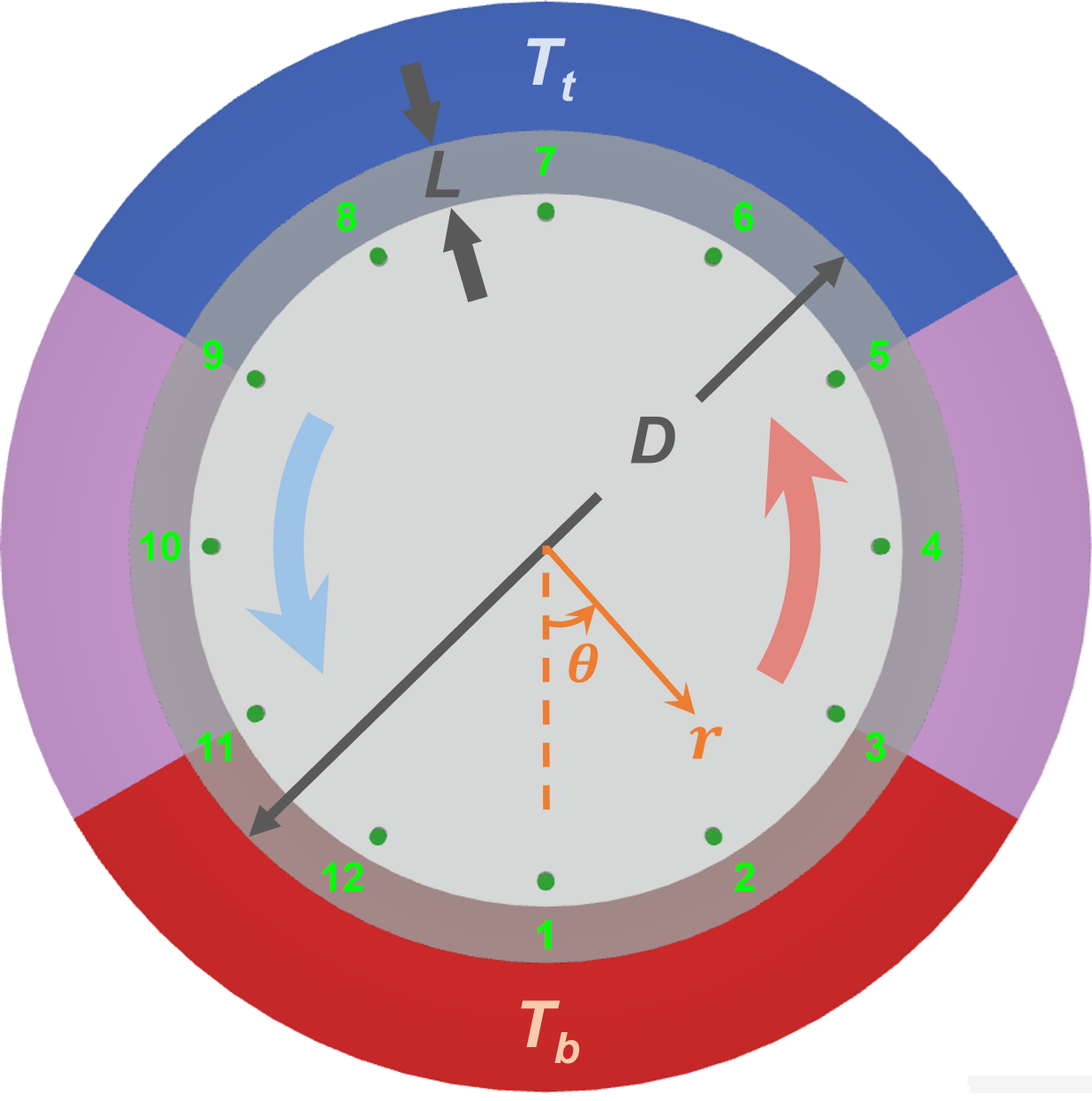}}
\caption{Sketch of the thin-disk convection cell.
The lower (red) and upper (blue) thirds of the circular periphery are held at temperatures $T_b$ and $T_t$, respectively; the two lateral sixths (purple) and front and back endwalls are Plexiglas.
Twelve thermistors (green) embedded in blind holes in the front endwall record the local temperature.
Arrows indicate a counter-clockwise LSC and its associated dipolar temperature field.
Polar coordinates are marked in orange.}
\label{fig1}
\end{figure}

Experimental details are reported elsewhere~\cite{wang2018}, and here we summarize the key features.
As Fig.~\ref{fig1} shows, the cell has diameter $D = 188$~mm and thickness $L = 20$~mm, with an aspect ratio $\Gamma\equiv L/D = 0.106$.
The upper and lower thirds of the circular sidewall are thermally conducting nickel-plated copper, while the remaining sidewall and the two flat endwalls are insulating Plexiglas.
Two parallel silicone-rubber film heaters provide uniform heating to the bottom copper plate at $T_b$, while a temperature-controlled chiller (NESLAB RTE740) maintains the top plate temperature at $T_t$ via circulating water. Each plate temperature is monitored by two calibrated thermistors at 2~Hz with 5~mK accuracy.
The control parameters are $Ra=\alpha g\Delta T D^3/(\nu\kappa)$ and Prandtl number $Pr=\nu/\kappa$, where $T=T_b-T_t$, $g$ is gravitational acceleration, and $\alpha$, $\nu$, and $\kappa$ are the fluid's thermal expansion coefficient, kinematic viscosity, and thermal diffusivity.
We used distilled water ($Pr=5.7$), a 10~wt.\% glycerol--water solution (GAS, $Pr=5.7$), and a 20~wt.\% GAS ($Pr=7.6$), spanning $2.0\times10^9\lesssim Ra\lesssim1.3\times10^{10}$. 
In this range the flow is turbulent and is dominated by a single LSC occupying the bulk~\cite{song2011a,wang2016,wang2018}.
Twelve equally spaced thermistors lie at radius 71~mm, 0.5~mm from the fluid-facing endwall surface, which sample the local temperature at approximately 1~Hz over about one month per $Ra$.

In our system, the thin quasi-2D geometry suppresses 3D secondary modes, including the torsional and sloshing motions observed in upright cylindrical cells~\cite{funfschilling2004,xi2009}, while the circular cross section eliminates the corner flows characteristic of rectangular cells~\cite{sugiyama2010}.
The system therefore isolates the essential ingredients of turbulent RBC: the LSC, viscous and thermal boundary layers (BLs), thermal plumes, and turbulent fluctuations.
Despite this simplified geometry, the LSC retains the key dynamics observed in more complex systems, including quasi-periodic oscillations, cessations, reversals, and small-scale fluctuations~\cite{song2011a,wang2018}, making it well suited for identifying their underlying physical mechanisms.

For $Pr\gtrsim1$, velocity measurements show that the bulk LSC is well approximated by a solid-body rotation~\cite{song2014},
\begin{equation}
    \mathbf{u}(r,\theta,t)=r\Omega(t)\mathbf{e}_\theta,
    \label{eq:velocity_ansatz}
\end{equation}
where $\Omega(t)$ is the instantaneous angular velocity.
Similar solid-body LSCs have been observed in upright cylindrical cells~\cite{qiu2001,sun2005}.
The corresponding bulk temperature field is dominated by a dipolar mode whose amplitude grows linearly with radius~\cite{qiu2001,brown2007,wang2018},
\begin{equation}
    T(r,\theta,t)=T_0(t)+\xi(t)\frac{r}{R}\cos[\theta-\phi(t)],
    \label{eq:temperature_ansatz}
\end{equation}
with $R=D/2$.
We fit Eq.~\eqref{eq:temperature_ansatz} to the temperature time series from the 12 sidewall thermistors, with the dipole amplitude $\xi(t)$, orientation $\phi(t)$, and azimuthally uniform component $T_0(t)$ as fitting parameters.

\begin{figure}
\centerline{\includegraphics[width=0.45\textwidth]{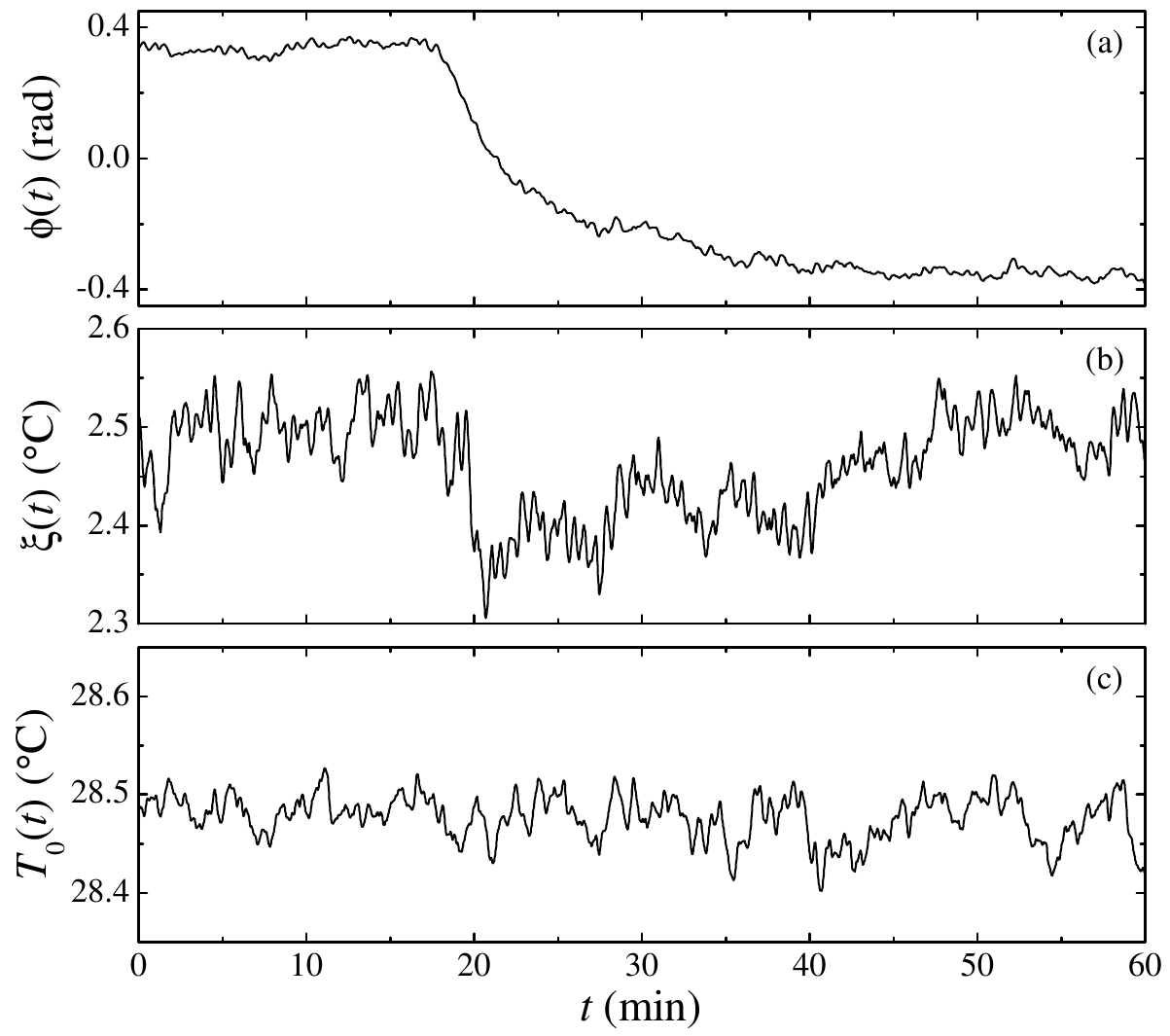}}
\caption{Measured time series of the amplitude $\phi(t)$~(a), orientation $\xi(t)$~(b), and azimuthally uniform component $T_0(t)$~(c) of the dipolar temperature field, obtained by fitting Eq.~\eqref{eq:temperature_ansatz} to the 12 thermistors. Data are for water at $Ra=7.4\times10^9$ and $Pr=5.7$.}
\label{fig2}
\end{figure}

Figure~\ref{fig2} shows a representative one-hour record.
For $t\lesssim18$~min, the system remains in a statistically steady LSC state, which occupies more than 90\% of the measurement time; $\xi(t)$ and $\phi(t)$ fluctuate about finite means with clear oscillatory components.
The nonzero $\phi$ breaks the left--right symmetry imposed by the boundary heating and cooling, thereby generating a net buoyancy torque that sustains the LSC [see Eq.~\eqref{eq:momentum}].
For $18\lesssim t\lesssim20$~min, a strong plume eruption disrupts the LSC~\cite{wang2018}, causing both $\xi$ and $\phi$ to collapse and restoring approximate left--right symmetry.
The LSC subsequently reforms with either circulation direction.
In the event shown, it reverses, accompanied by a sign change in $\phi(t)$.
Throughout the process, $T_0(t)$ fluctuates about its mean without a discernible coherent oscillation, supporting the separation between the dipolar LSC mode and the approximately isotropic turbulent background.

We next derive a low-dimensional stochastic model for the LSC dynamics.
Under the Boussinesq approximation, the flow obeys the incompressible Navier--Stokes and temperature equations,
\begin{gather}
    \frac{\partial \mathbf{u}}{\partial t}
    + \mathbf{u}\cdot\nabla \mathbf{u}
    = -\frac{1}{\rho}\nabla p
    + \nu \nabla^2 \mathbf{u}
    - \alpha g (T-T_0)\mathbf{e}_g, \label{eq:NS}\\
    \frac{\partial (T-T_0)}{\partial t}
    + \mathbf{u}\cdot\nabla (T-T_0)
    = \kappa \nabla^2 (T-T_0), \label{eq:temperature_equation} \\
    \nabla\cdot\mathbf{u}=0, \label{eq:continuity}
\end{gather}
where $\mathbf{e}_g$ is the upward unit vector.
Because $T_0(t)$ is azimuthally uniform, it vanishes upon projection of the temperature equation onto the dipolar harmonic.
Substituting Eqs.~\eqref{eq:velocity_ansatz} and \eqref{eq:temperature_ansatz} and averaging over the cell eliminates the bulk Laplacian terms.
We therefore model viscous and thermal diffusion through the BLs using the Prandtl--Blasius-type thickness estimates $\lambda_u\sim(\nu/|\Omega|)^{1/2}$ and $\lambda_T\sim Pr^{-1/3}\lambda_u$ for the viscous and thermal BLs, respectively\cite{shishkina2010}.
This yields three dimensionless equations [see Supplemental Material (SM) for the derivation]:
\begin{align}
    &\dot{\tilde{\Omega}}
    =
    -a\,\tilde{\Omega}\sqrt{|\tilde{\Omega}|}
    + \,\tilde{\xi}\sin\phi + \eta(t), \label{eq:momentum}\\
    &\dot{\tilde{\xi}}
    =
    b\,\sqrt{|\tilde{\Omega}|}
    \left(
        c\cos\phi - \tilde{\xi}
    \right), \label{eq:amplitude}\\
    &\dot{\phi}
    =
    \tilde{\Omega} - bc\,\sqrt{|\tilde{\Omega}|}\frac{\sin\phi}{\tilde{\xi}}, \label{eq:angle}
\end{align}
where $\tilde{\Omega}=\Omega t_f$, $\tilde{\xi}=\xi/\Delta T$, and $t_f=\sqrt{D/(\alpha g\Delta T)}$ is the free-fall time.
We use white noise $\eta(t)$ as a minimal effective representation of unresolved turbulent forcing, with $\langle\eta(t)\eta(t')\rangle=2W\delta(t-t')$, where $W$ is the noise intensity and $\langle\cdots\rangle$ denotes a time average.
To leading order, the noise is included only in the momentum equation because the fluctuations are assumed to act through buoyancy.
In Eqs.~\eqref{eq:momentum}--\eqref{eq:angle}, the positive dimensionless parameters are
\begin{align}
    &a = \left(6+\frac{3}{\Gamma}\right)Ra^{-1/4}Pr^{1/4}, \label{eq:a} \\
    &b = 6N Ra^{-1/4} Pr^{-5/12}, \label{eq:b} \\
    &c = \frac{\sqrt{3}}{\pi N} \label{eq:c}.
\end{align}
Here $N\geq1$ parametrizes additional relaxation of the dipolar temperature mode by thermal plumes, with $N=1$ recovering the boundary-layer-only estimate.
The momentum Eq.~\eqref{eq:momentum} resembles the BL drag--buoyancy balance as an earlier LSC model~\cite{brown2007}, whereas Eqs.~\eqref{eq:amplitude} and \eqref{eq:angle} dynamically close the temperature amplitude and orientation.

\begin{figure}
\centerline{\includegraphics[width=0.42\textwidth]{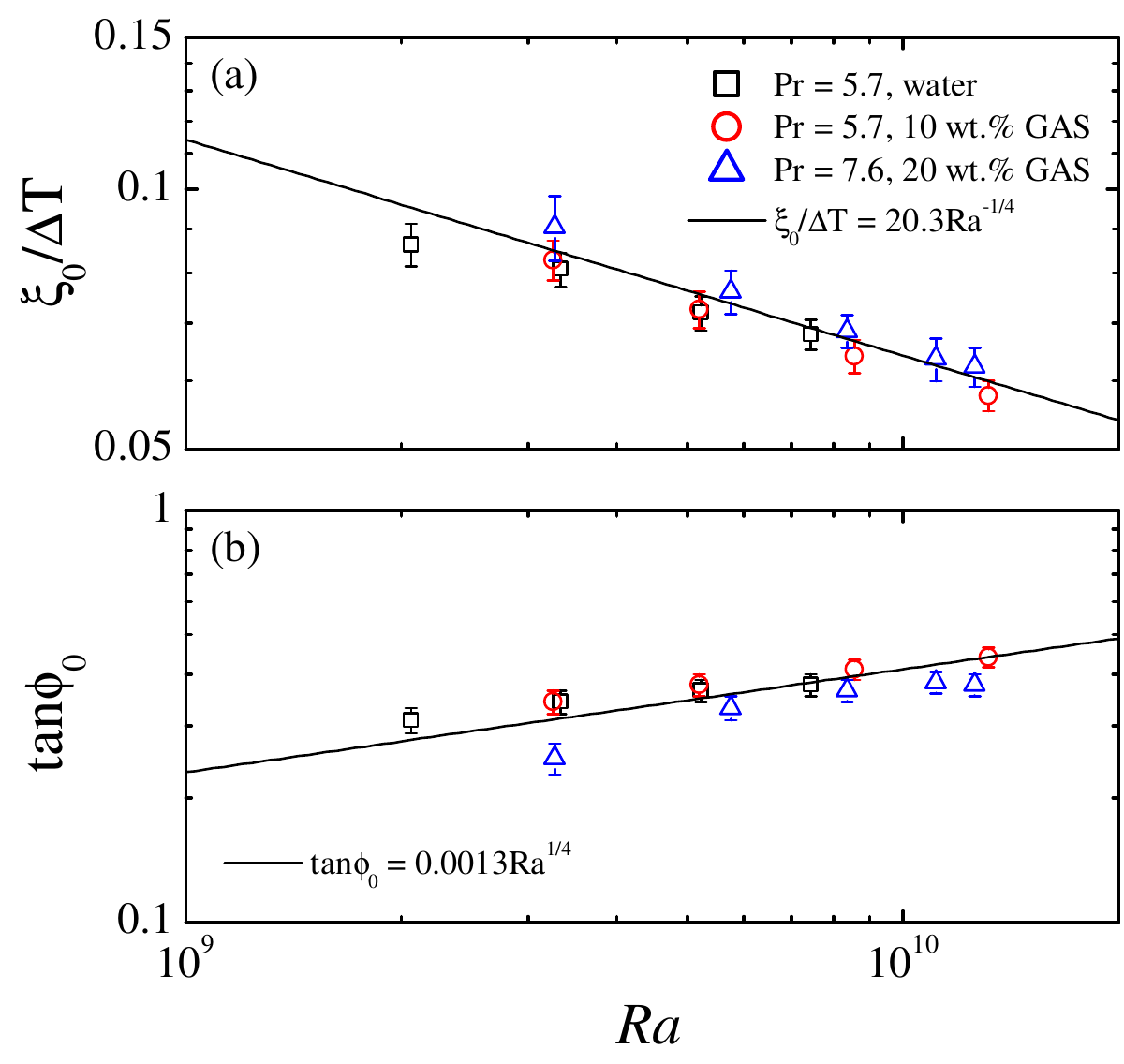}}
\caption{Time-averaged dipolar amplitude $\langle\xi\rangle/\Delta T$ (a) and orientation $\tan\phi_0$ (b) in statistically steady LSC intervals, for the three working fluids. Solid lines show the measured power-law trends $Ra^{-1/4}$ and $Ra^{1/4}$, respectively.}
\label{fig3}
\end{figure}

The deterministic parts of Eqs.~\eqref{eq:momentum}--\eqref{eq:angle} admit a continuum of no-LSC equilibria, $(\tilde{\Omega},\tilde{\xi},\phi)=(0,\tilde{\xi}^*,0)$ with $\tilde{\xi}^*>0$.
While the vector field is nonanalytic at $\tilde{\Omega}=0$, leading-order perturbation analysis yields an unstable branch $\propto\exp(\sqrt{\tilde{\xi}^*}\,t)$.
Thus the no-LSC state is unstable, and the LSC emerges through spontaneous symmetry breaking, consistent with the experimentally observed reformation of the LSC after collapse.

There are another two symmetry-related equilibria $(\pm\tilde{\Omega}_0,\tilde{\xi}_0,\pm\phi_0)$, which corresponding to finite LSCs with opposite rotation directions.
The fixed-point relations at asymptotically large $Ra$ are
\begin{align}
    \tilde{\Omega}_0
    &\simeq
    \sqrt{bc/a} \simeq 0.31Pr^{-1/3}, \label{eq:omega_0} \\
    \tilde{\xi}_0
    &\simeq
    a^{1/4}\left(bc\right)^{3/4} \simeq5.93Ra^{-1/4} Pr^{-1/4}, \label{eq:xi_0} \\
    \tan\phi_0
    &\simeq
    \left(\frac{c}{ab^3}\right)^{1/4} \simeq \frac{0.093}{N}Ra^{1/4} Pr^{1/4}. \label{eq:phi_0}
\end{align}
The corresponding Reynolds number $Re=\tilde{\Omega}_0RD/(t_f\nu)$ has the asymptotic scaling
\begin{equation}
    Re \simeq 0.16Ra^{1/2}Pr^{-5/6},
    \label{eq:Reynolds}
\end{equation}
which agrees well with the measured $Ra$ dependence of the mean circulation speed~\cite{song2011a} and with the BL-dominated branch of the Grossmann--Lohse scaling theory~\cite{grossmann2000}.
As shown in Fig.~\ref{fig3}, the measured $Ra$ exponents of $\langle\xi\rangle/\Delta T$ and $\tan\phi_0$ agree with Eqs.~\eqref{eq:xi_0} and \eqref{eq:phi_0}.
The prefactors agree only semiquantitatively: for $Pr=7.6$, the measured amplitude is about five times the leading-order prediction, while fitting $\tan\phi_0$ gives $N\simeq118.7$, far above the $N\simeq9.8$ inferred from mode damping (Fig.~\ref{fig5}).
These discrepancies likely reflect the simplified transport closure, in particular the omission of order-unity prefactors in the BL-thickness estimates and the truncation of the piecewise thermal boundary forcing to its first azimuthal harmonic.
The available $Pr$ range is too narrow to resolve the predicted $Pr$ dependences.

\begin{figure}
\centerline{\includegraphics[width=0.4\textwidth]{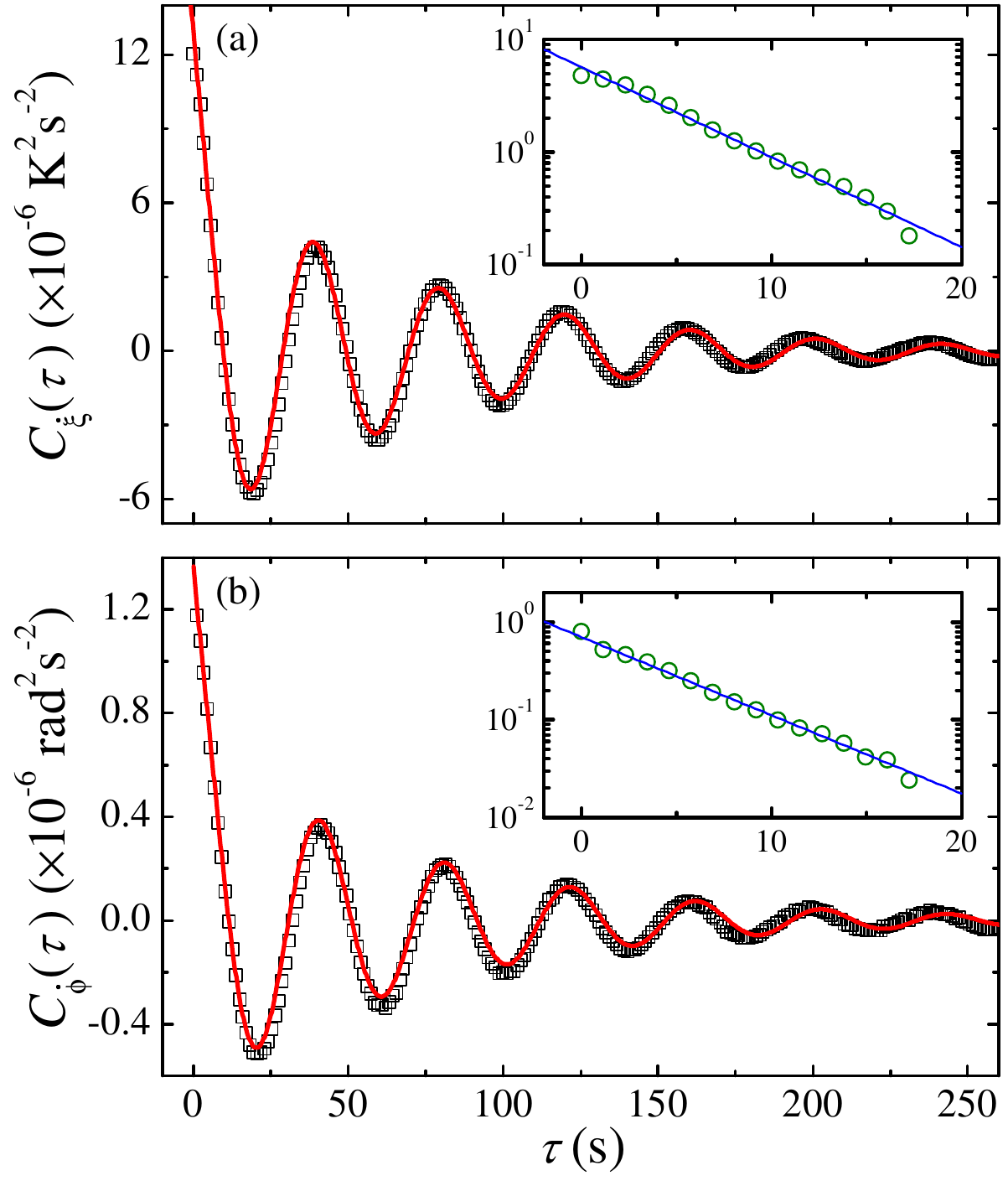}}
\caption{Measured autocorrelation functions of $\dot\xi(t)$~(a) and $\dot\phi(t)$~(b) for water at $Ra=7.4\times10^9$ and $Pr=5.7$ during statistically steady LSC intervals.
Red curves are fits to Eq.~\eqref{eq:VACF} with $\gamma_1=0.184~\mathrm{s}^{-1}$, $\gamma_2=0.0135~\mathrm{s}^{-1}$, and $\omega_0=0.155~\mathrm{s}^{-1}$.
Insets show the residual after subtracting the fitted oscillatory contribution, where blue curves show the fitted purely exponential component.}
\label{fig4}
\end{figure}

We next analyze linear perturbations about the finite-LSC state. Over the experimental parameter range, the Jacobian has one real eigenvalue and a complex-conjugate pair, which at sufficiently large $Ra$ are
\begin{align}
    \sigma_1
    &\simeq
    -2a\sqrt{\tilde{\Omega}_0}
    \simeq -38.1Ra^{-1/4}Pr^{1/12}, \label{eq:gamma1}\\
    \sigma_{2,3}
    &=\sigma_r \pm i\sigma_i
    \simeq
    \left(\frac{a}{4}-b\right)
    \sqrt{\tilde{\Omega}_0}
    \pm i\tilde{\Omega}_0. \label{eq:complex_mode}
\end{align}
Thus the finite-LSC states are stable when $b>a/4$. The real eigenmode relaxes without oscillation, whereas the complex pair describes a damped oscillation whose leading frequency is the dimensionless LSC turnover rate.
At $Pr=7.6$, Eq.~\eqref{eq:complex_mode} becomes
\begin{equation}
    \sigma_r\simeq(5.65-1.03N)Ra^{-1/4}, \quad \sigma_i\simeq0.158.
    \label{eq:gamma2}
\end{equation}
For a linear stochastic system, correlations near a fixed point inherit the same eigenmodes.
We therefore examine the autocorrelation funcstions (ACFs) of $\dot{\xi}$ and $\dot{\phi}$, $C_{\dot{\xi}}(\tau)=\langle\dot{\xi}(t+\tau)\dot{\xi}(t)\rangle$ and $C_{\dot{\phi}}(\tau)=\langle\dot{\phi}(t+\tau)\dot{\phi}(t)\rangle$, which suppress slow LSC wandering while retaining the local decay rates and oscillation frequency. Figure~\ref{fig4} shows that both are well described by
\begin{equation}
    C_{\dot{\xi},\dot{\phi}}(\tau)
    =
    A_1e^{-\gamma_1\tau}
    +A_2e^{-\gamma_2\tau}\cos\left(\omega_0\tau+\varphi_0\right).
    \label{eq:VACF}
\end{equation}
The two ACFs require different amplitudes and phases but share, within experimental uncertainty, the same $\gamma_1$, $\gamma_2$, and $\omega_0$, as expected if they project onto the same three eigenmodes.

\begin{figure}
\centerline{\includegraphics[width=0.4\textwidth]{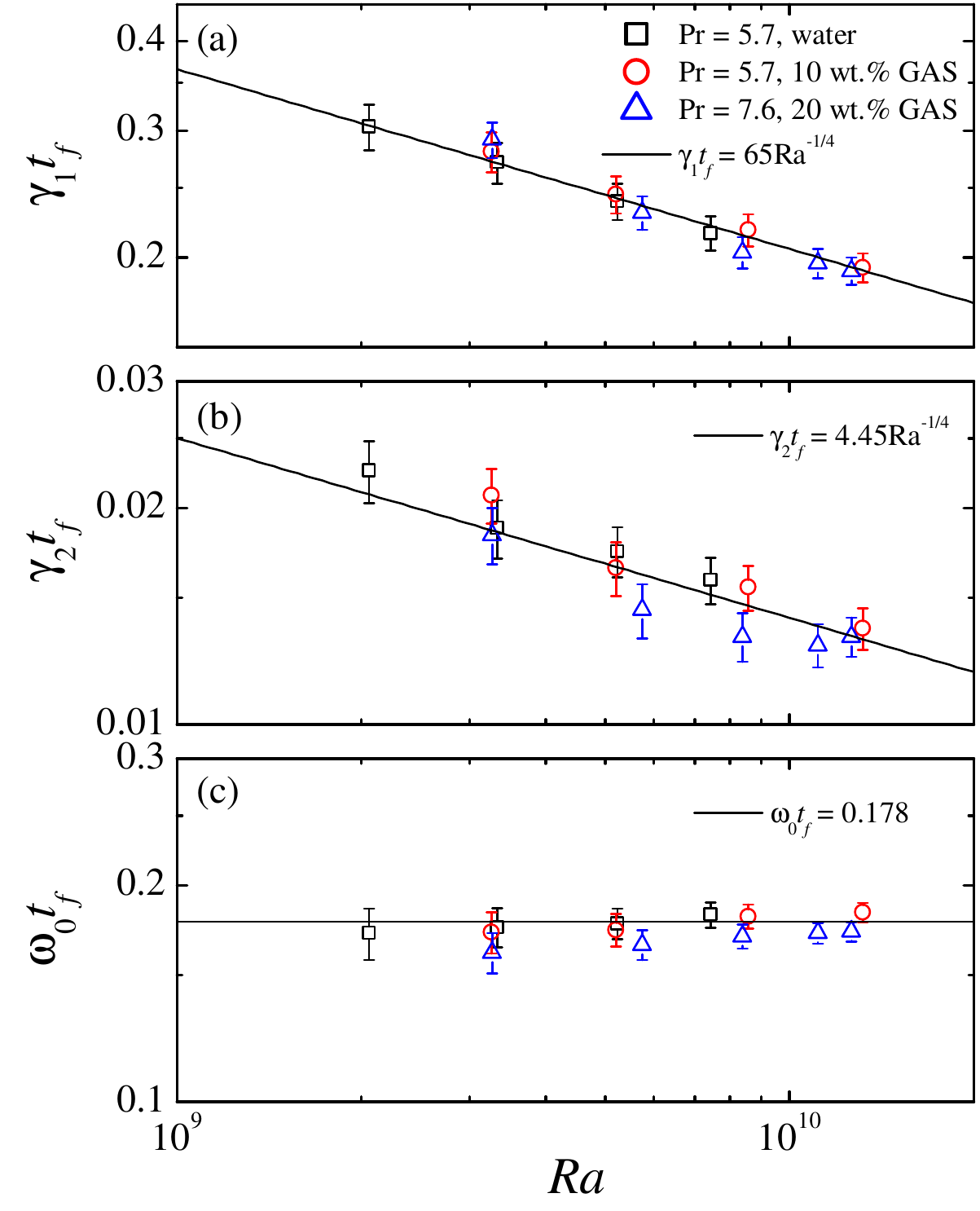}}
\caption{Measured normalized decay rates $\gamma_1t_f$~(a) and $\gamma_2t_f$~(b), and oscillation frequency $\omega_0t_f$~(c), obtained from autocorrelation functions for the three working fluids.
Solid lines are power-law fits using the exponents predicted by Eq.~\eqref{eq:gamma1} and the real and imaginary parts of Eq.~\eqref{eq:gamma2}, with $N\simeq9.8$.}
\label{fig5}
\end{figure}

Figure~\ref{fig5} shows that $\gamma_1t_f$ and $\gamma_2t_f$ scale approximately as $Ra^{-1/4}$, while $\omega_0t_f$ is nearly independent of $Ra$, consistent with the leading exponents in Eqs.~\eqref{eq:gamma1} and \eqref{eq:gamma2}.
Fitting $\gamma_2t_f$ gives $N\simeq9.8$, indicating substantial plume-enhanced relaxation of the dipolar temperature mode, while the fitted prefactor of $\gamma_1t_f$ is 1.4 times the model prediction at $Pr=7.6$.
The prefactor differences are consistent with the same simplified BL and thermal-mode approximations discussed above.
The robust result is therefore the predicted modal structure and scaling: one fast relaxational mode and a weakly damped oscillatory pair whose frequency is set by the LSC turnover rate.
Their direct observation supports the model’s central physical picture and, to our knowledge, provides the first minimal framework to predict LSC stability eigenmodes and quantitatively test their decay rates and oscillation frequencies against experiment.

The physical feedback underlying the oscillatory pair is transparent in Eqs.~\eqref{eq:momentum}--\eqref{eq:angle}. Advection by the LSC rotates the thermal dipole away from the symmetry axis of the imposed heating and cooling; the resulting transverse temperature component supplies the buoyancy torque that maintains the LSC.
Thermal relaxation, in turn, drives the dipole back toward the imposed symmetry axis.
Linearization of this coupled torque--relaxation loop produces the complex eigenvalues.
Because their real part is negative, persistent experimental oscillations require continual excitation by turbulent fluctuations and therefore appear as broadened quasi-periodic spectral peaks rather than a self-sustained limit cycle.
The same stochastic equations also provide a route to cessations and reversals: sufficiently large fluctuations can drive trajectories from a finite-LSC attractor into the neighborhood of the unstable no-LSC set, after which either LSC branch can be re-established. This dynamical picture is consistent with the plume-triggered collapses observed experimentally~\cite{wang2018}, while a quantitative rare-event theory for the transition rates is beyond the present closure.

In summary, projecting the Boussinesq momentum and temperature equations onto the dominant circulation and dipolar thermal modes yields a three-variable stochastic model for turbulent RBC.
The model produces two symmetry-related circulating states and predicts, about each state, one purely relaxational mode and a damped oscillatory pair whose frequency is set by the turnover rate.
Measurements in a thin-disk cell resolve this modal structure and reproduce its principal $Ra$ scalings.
The framework thereby separates three aspects of LSC dynamics---mean-state selection, intrinsic linear modes, and fluctuation-driven transitions---within a common reduced description.

Digital data associated with this work are available from DataSpace at Princeton University~\cite{data}.

This research was supported by U.S. DOE (Contract No. DE-AC02-09CH11466).


\bibliography{references}

\end{document}


\preprint{APS/123-QED}

\title{Stable eigenmodes of the Large-scale circulation in Turbulent Convection: Supplementary Materials}

\author{Yin Wang$^1$}
 \email{ywang3@pppl.gov}
\author{Pik-Yin Lai$^2$}
\author{Penger Tong$^3$}
\affiliation{$^1$Princeton Plasma Physics Laboratory, Princeton University, Princeton, New Jersey 08543, USA \\
             $^2$Department of Physics and Center for Complex Systems, National Central University, Chungli, Taiwan 320, R.O.C. \\
             $^3$Department of Physics, Hong Kong University of Science and Technology, Hong Kong}%



\date{\today}

\begin{abstract}

\end{abstract}

\maketitle

\section{Derivation of the model for LSC dynamics}

We begin with the Boussinesq Navier--Stokes and temperature equations an incompressible fluid,
\begin{gather}
    \frac{\partial \mathbf{u}}{\partial t}
    + \mathbf{u}\cdot\nabla \mathbf{u}
    = -\frac{1}{\rho}\nabla p
    + \nu \nabla^2 \mathbf{u}
    + \alpha g (T-T_0)\mathbf{e}_g, \label{eq:NS} \\
    \frac{\partial (T-T_0)}{\partial t}
    + \mathbf{u}\cdot\nabla (T-T_0)
    = \kappa \nabla^2 (T-T_0), \label{eq:temperature} \\
    \nabla\cdot\mathbf{u}=0, \label{eq:continuity}
\end{gather}
where $\mathbf{e}_g$ is the unit vector in the vertically upward direction.
Because $T_0(t)$ is the azimuthally uniform temperature component, its time derivative drops out when the temperature equation is projected onto the dipolar mode considered below.
Owing to the quasi-two-dimensional (2D) geometry, we use the polar coordinates
$(r,\theta)$ defined in Fig.~1 of the main text.
The bulk LSC is approximated as solid-body rotation,
\begin{equation}
    \mathbf{u} = \Omega(t)\, r\, \mathbf{e}_\theta,
    \label{eq:ansatz_u}
\end{equation}
which satisfies Eq.~\eqref{eq:continuity}.
The associated bulk temperature field is approximated by its first azimuthal harmonic,
\begin{equation}
    T(r,\theta,t)=T_0+\xi(t)\frac{r}{R}\cos\!\left[\theta-\phi(t)\right],
    \label{eq:ansatz_T}
\end{equation}
where $R=D/2$ is the cell radius.
For these bulk fields,
\begin{equation}
    \nabla^2 \mathbf{u}=0,
    \qquad
    \nabla^2(T-T_0)=0,
\label{eq7}
\end{equation}
so viscous and thermal diffusion enter through the boundary layers (BLs).

\subsection{Momentum equation}

Projecting Eq.~\eqref{eq:NS} onto the azimuthal direction and averaging over the disk area gives, for the unsteady term,
\begin{equation}
    \left\langle \frac{\partial u_\theta}{\partial t}\right\rangle_A
    =
    \frac{1}{\pi R^2}\int_0^R\int_0^{2\pi}\dot\Omega r \, r\, d\theta\,dr
    =
    \frac{2R}{3}\dot\Omega,
    \label{eq:momentum_time_derivative}
\end{equation}
where $\langle\cdots\rangle_A$ denotes an area average.
The azimuthal component of the nonlinear advection term vanishes for solid-body rotation,
\begin{equation}
    \big[(\mathbf{u}\cdot\nabla)\mathbf{u}\big]_\theta
    =
    u_r\partial_r u_\theta
    + \frac{u_\theta}{r}\partial_\theta u_\theta
    + \frac{u_r u_\theta}{r}=0,
    \label{eq:momentum_advection}
\end{equation}
and the azimuthally averaged pressure-gradient term also vanishes,
\begin{equation}
    \left\langle \frac{1}{r}\frac{\partial p}{\partial \theta}\right\rangle_A
    =
    \frac{1}{\pi R^2}\int_0^Rdr\int_0^{2\pi}\frac{\partial p}{\partial\theta}\,d\theta
    =0.
\label{eq:momentum_pressure}
\end{equation}

We model viscous diffusion using a Prandtl--Blasius BL thickness~\cite{shishkina2010}
\begin{equation}
    \lambda_u \sim \sqrt{\frac{\nu}{|\Omega|}}
\label{eq:viscous_BL}
\end{equation}
with $\lambda_u \ll R,L$, where $L$ is the cell thickness.
Within the viscous BL, the azimuthal viscous diffusion is approximated as $\nu[\nabla^2\mathbf{u}]_\theta\simeq-\nu\Omega R/\lambda_u^2$.
Accounting for viscous BLs along the circular periphery and the front and back endwalls, the total viscous contribution is
\begin{equation}
    \nu\left\langle
    \big[\nabla^2\mathbf{u}\big]_\theta
    \right\rangle_V
    \simeq
    -\nu\frac{\Omega R}{\lambda_u^2}\,
    \frac{2\pi R^2\lambda_u+2\pi RL\lambda_u}{\pi R^2L}
    =
    -\left(2+\frac{1}{\Gamma}\right)\sqrt{\nu}\Omega\sqrt{|\Omega|},
\label{eq:momentum_diffusion}
\end{equation}
where $\Gamma=L/D$ is the aspect ratio of the convection cell.
In Eq.\eqref{eq:momentum_diffusion}, the $1/\Gamma$ term accounts for additional viscous diffusion from the BLs at the front and back endwalls; all other terms are captured by the 2D polar-coordinate description.

The azimuthal buoyancy contribution is
\begin{equation}
\begin{aligned}
    \alpha g\left\langle \big[(T-T_0)\mathbf{e}_g\big]_\theta\right\rangle_A
    =
    \frac{\alpha g}{\pi R^2}
    \int_0^R\int_0^{2\pi}
    \xi\frac{r}{R}\cos(\theta-\phi)\sin\theta\, rd\theta dr 
    =
    \frac{\alpha g}{3}\xi\sin\phi.
\end{aligned}
\label{eq:momentum_buoyancy}
\end{equation}
Combining Eqs.~\eqref{eq:momentum_time_derivative}--\eqref{eq:momentum_buoyancy} yields
\begin{equation}
    \dot\Omega
    =
    -\left(3 + \frac{3}{2\Gamma} \right)
    \frac{\sqrt{\nu}}{R}\Omega\sqrt{|\Omega|}
    + \frac{\alpha g}{2R}\,\xi\sin\phi.
\label{eq:momentum}
\end{equation}
The derivation of Eq.~\eqref{eq:momentum} follows the same approach as the LSC amplitude equation in Ref.~\cite{brown2007}.

\subsection{Temperature equations}

We next project the temperature equation onto the dipolar mode.
Radially averaging the unsteady term and advection term gives
\begin{gather}
    \left\langle \frac{\partial (T-T_0)}{\partial t}\right\rangle_r
    =
    \frac{2}{R^2}\int_0^R
    \frac{r}{R} \frac{\partial}{\partial t}
    \bigg[\xi(t)\cos\!\left[\theta-\phi(t)\right]\bigg]\, rdr
    =
    \frac{2}{3}
    \left(
        \dot\xi\cos\phi-\xi\dot\phi\sin\phi
    \right)\cos\theta
    +
    \frac{2}{3}
    \left(
        \dot\xi\sin\phi+\xi\dot\phi\cos\phi
    \right)\sin\theta, \label{eq:thermal_time_derivative} \\
    \left\langle \mathbf{u}\cdot\nabla (T-T_0)\right\rangle_r
    =
    \frac{2}{R^2}\int_0^R
    \Omega \frac{\partial (T-T_0)}{\partial\theta}\, r\,dr
    =
    \frac{2\Omega\xi}{3}
    \left[
        \sin\phi\cos\theta-\cos\phi\sin\theta
    \right], \label{eq:thermal_advection}
\end{gather}
where $\langle f\rangle_r=(2/R^2)\int_0^R f\,rdr$.
Thermal diffusion is modeled using the corresponding laminar BL estimate~\cite{shishkina2010},
\begin{equation}
    \lambda_T \simeq Pr^{-1/3}\sqrt{\frac{\nu}{|\Omega|}}.
    \label{eq:thermal_BL}
\end{equation}
The first azimuthal harmonic of the imposed constant-temperature heating and cooling boundary profile is
\begin{equation}
    \frac{\sqrt{3}\Delta T}{\pi}\cos\theta,
\label{eq:boundary_forcing}
\end{equation}
where the dipolar coefficient $\sqrt{3}\Delta T/\pi$ is obtained from the first Fourier projection of the constant-temperature boundary heating and cooling profile,
\begin{equation}
    \frac{1}{\pi}
    \int_{0}^{2\pi}
    \left(
    T_{\rm wall}(\theta)-T_0\right)
    \cos\theta\,d\theta
    =
    \frac{\sqrt{3}\Delta T}{\pi},
    \label{eq:boundary_forcing_detail}
\end{equation}
where the hot and cold conducting arcs each span $2\pi/3$.
Within the thermal BL, the radially averaged diffusive term is therefore approximated as
\begin{equation}
    \kappa\left\langle \nabla^2(T-T_0)\right\rangle_r
    \simeq\frac{2\kappa}{R\lambda_T}
    \left[
     \frac{\sqrt{3}\Delta T}{\pi}\cos\theta
    -
    N\xi\cos(\theta-\phi)
    \right]
    =\frac{2\sqrt{\kappa}}{R}
    Pr^{-1/6}\sqrt{|\Omega|}
    \left[
    \frac{\sqrt{3}\Delta T}{\pi}\cos\theta
    -
    N\xi\cos(\theta-\phi)
    \right].
    \label{eq:thermal_diffusion}
\end{equation}
Here $N\ge1$ is a phenomenological enhancement factor that accounts for additional relaxation of the bulk dipolar temperature mode by plume-mediated transport.
The wall forcing itself remains fixed by the imposed boundary temperature profile.

Combining Eqs.~\eqref{eq:thermal_time_derivative}, \eqref{eq:thermal_advection}, and \eqref{eq:thermal_diffusion} and collecting the coefficients of $\cos\theta$ and $\sin\theta$ gives
\begin{align}
    &\dot\xi\cos\phi-\xi\dot\phi\sin\phi
    + \Omega\xi\sin\phi
    + \frac{3\sqrt{\kappa}}{R} Pr^{-1/6}\sqrt{|\Omega|}
    \left(
        N\xi\cos\phi - \frac{\sqrt{3}\Delta T}{\pi}
    \right)=0, \label{eq:thermal_cos} \\
    &\dot\xi\sin\phi+\xi\dot\phi\cos\phi
    - \Omega\xi\cos\phi
    + \frac{3\sqrt{\kappa}}{R} Pr^{-1/6}\sqrt{|\Omega|}
    N\xi\sin\phi=0. \label{eq:thermal_sin}
\end{align}
Because $\cos\theta$ and $\sin\theta$ are linearly independent, Eqs.~\eqref{eq:thermal_cos} and \eqref{eq:thermal_sin} must hold separately.
Solving these two equations for $\dot\xi$ and $\dot\phi$ yields
\begin{align}
    \dot\xi
    &=
    \frac{3\sqrt{\kappa}}{R} Pr^{-1/6}\sqrt{|\Omega|}
    \left[
        \frac{\sqrt{3}\Delta T}{\pi}\cos\phi -  N\xi
    \right],    \label{eq:amplitude}\\
    \dot\phi
    &=
    \Omega
    - \frac{3\sqrt{3\kappa}\Delta T}{\pi R} Pr^{-1/6}\sqrt{|\Omega|}\,\frac{\sin\phi}{\xi}. \label{eq:angle}
\end{align}
Equations~\eqref{eq:momentum}, \eqref{eq:amplitude}, and \eqref{eq:angle} form a closed deterministic model for the coupled evolution of the LSC angular velocity, thermal dipole amplitude, and orientation.

\subsection{Dimensionless form}

Using the free-fall time, $t_f=\sqrt{2R/\alpha g\Delta T}$, we define
$\tilde{\Omega}=\Omega t_f$, $\tilde{\xi}=\xi/\Delta T$, and $\tilde{t}=t/t_f$.
Relabeling $\tilde{t}$ as $t$ for brevity, Eqs.~\eqref{eq:momentum}, \eqref{eq:amplitude}, and \eqref{eq:angle} can be rewritten in to dimensionless forms, with 
\begin{align}
    \dot{\tilde{\Omega}}
    &=
    -a\,\tilde{\Omega}\sqrt{|\tilde{\Omega}|}
    + \,\tilde{\xi}\sin\phi, \label{eq:momentum_normalized}\\
    \dot{\tilde{\xi}}
    &= 
    b\,\sqrt{|\tilde{\Omega}|}
    \left(
        c\cos\phi - \tilde{\xi}
    \right), \label{eq:amplitude_normalized}\\
    \dot{\phi}
    &= 
    \tilde{\Omega} - bc\,\sqrt{|\tilde{\Omega}|}\, \frac{\sin\phi}{\tilde{\xi}}, \label{eq:angle_normalized}
\end{align}
with
\begin{equation}
    a = \left(6+\frac{3}{\Gamma}\right)Ra^{-1/4}Pr^{1/4}, \quad
    b = 6N Ra^{-1/4} Pr^{-5/12}, \quad
    c = \frac{\sqrt{3}}{\pi N}
    \label{eq:abc}
\end{equation}
Adding a white-noise forcing $\eta(t)$ to Eq.~\eqref{eq:momentum_normalized}, with
$\langle\eta(t)\eta(t')\rangle=2W\delta(t-t')$, gives the Eqs.~(6)--(11) in the main text.

\section{Linear stability analysis}

Introducing variables $X=\tilde{\xi}\cos\phi$ and $Y=\tilde{\xi}\sin\phi$, Eqs.~\eqref{eq:momentum_normalized}--\eqref{eq:angle_normalized} can be recast as
\begin{align}
    \dot{\tilde{\Omega}} =
    & f_1(\tilde{\Omega},X,Y)= -a\tilde{\Omega}\sqrt{|\tilde{\Omega}|}
    + Y, \label{eq:momentum_XY}\\
    \dot{X} =
    & f_2(\tilde{\Omega},X,Y) = -\tilde{\Omega}Y
    + b\,\sqrt{|\tilde{\Omega}|}(c-X), \label{eq:X}\\
    \dot{Y} =
    & f_3(\tilde{\Omega},X,Y) = \tilde{\Omega}X
    - b\,\sqrt{|\tilde{\Omega}|}Y. \label{eq:Y}
\end{align}
This representation avoids the numerically problematic ratio
$\sin\phi/\tilde{\xi}$ when $\tilde{\xi}$ approaches zero.

For $\tilde{\Omega}\neq0$, the Jacobian matrix of Eqs.~\eqref{eq:momentum_XY}--\eqref{eq:Y} is
\begin{equation}
    J=
    \begin{pmatrix}
        \dfrac{\partial f_1}{\partial \tilde{\Omega}}
        & \dfrac{\partial f_1}{\partial X}
        & \dfrac{\partial f_1}{\partial Y} \\
        \dfrac{\partial f_2}{\partial \tilde{\Omega}}
        & \dfrac{\partial f_2}{\partial X}
        & \dfrac{\partial f_2}{\partial Y} \\
        \dfrac{\partial f_3}{\partial \tilde{\Omega}}
        & \dfrac{\partial f_3}{\partial X}
        & \dfrac{\partial f_3}{\partial Y}
    \end{pmatrix}
    =
    \begin{pmatrix}
        -\dfrac{3}{2}a\sqrt{|\tilde{\Omega}|}
        & 0
        & 1 \\
        - Y + \frac{1}{2}b(c-X)\operatorname{sgn}({\tilde{\Omega}})/\sqrt{|\tilde{\Omega}|}
        & - b\sqrt{|\tilde{\Omega}|}
        & - \tilde{\Omega}\\
        X - \frac{1}{2}bY\operatorname{sgn}({\tilde{\Omega}})/\sqrt{|\tilde{\Omega}|}
        & \tilde{\Omega}
        & - b\sqrt{|\tilde{\Omega}|}
    \end{pmatrix}
    \label{eq:jacobian}
\end{equation}
with $\operatorname{sgn}(\dots)$ being the sign function.

\subsection{No-LSC equilibrium set}

Equations~\eqref{eq:momentum_XY}--\eqref{eq:Y} admit a continuum of
no-LSC equilibria,
\begin{equation}
    (\tilde{\Omega},X,Y)=(0,X^*,0).
    \label{eq33}
\end{equation}
We focus on $X^*>0$, corresponding to a thermal dipole aligned with the imposed first harmonic.  Because the vector field is nonanalytic at $\tilde{\Omega}=0$, a conventional three-dimensional Jacobian linearization is not defined there.
The leading transverse dynamics can, however, be obtained directly.
We introduce infinitesimal perturbations of the form
\begin{equation}
    (\tilde{\Omega},X,Y)
    =
    \left(
    \tilde{\Omega}^{\prime},
    X^*,
    Y^{\prime}
    \right),
    \label{eq34}
\end{equation}
with
\begin{equation}
    \begin{pmatrix}
        \tilde{\Omega}^{\prime}\\
        Y^{\prime}
    \end{pmatrix}
    =
    \begin{pmatrix}
        \varepsilon_1\\
        \varepsilon_2
    \end{pmatrix}
    e^{-i\omega t},
\label{eq35}
\end{equation}
where $\varepsilon_1$ and $\varepsilon_2$ are infinitesimal amplitudes of the same asymptotic order. To leading order, Eqs.~\eqref{eq:momentum_XY} and
\eqref{eq:Y} reduce to
\begin{equation}
    \dot{\tilde{\Omega}}^{\prime}=Y^{\prime},
    \qquad
    \dot{Y}^{\prime}=X^*\tilde{\Omega}^{\prime}.
    \label{eq36}
\end{equation}
Substituting Eq.~\eqref{eq35} gives
\begin{equation}
    -i\omega\varepsilon_1=\varepsilon_2,
    \qquad
    -i\omega\varepsilon_2=X^*\varepsilon_1 .
    \label{eq37}
\end{equation}
Eliminating $\varepsilon_2$ yields
\begin{equation}
    \omega^2=-X^*,
    \label{eq38}
\end{equation}
and therefore
\begin{equation}
    \omega=\pm i\sqrt{X^*}.
    \label{eq39}
\end{equation}
With the convention $e^{-i\omega t}$, the root $\omega=+i\sqrt{X^*}$ corresponds to an exponentially growing
perturbation,
\begin{equation}
    \tilde{\Omega}^{\prime},Y^{\prime}
    \propto e^{\sqrt{X^*}t}.
    \label{eq40}
\end{equation}
Hence every no-LSC fixed point with $X^*>0$ is transversely unstable, with leading growth rate $\sqrt{X^*}$.

The perturbation in $X$ is generated simultaneously through Eq.~\eqref{eq:X}.
Writing $X=X^*+X^{\prime}$, its leading evolution is
\begin{equation}
    \dot{X}^{\prime}
    \simeq
    b(c-X^*)\sqrt{|\tilde{\Omega}^{\prime}|}.
    \label{eq41}
\end{equation}
Although this term is nonanalytic and scales as $|\tilde{\Omega}^{\prime}|^{1/2}$, $X^{\prime}$ enters the $Y$ equation only through the product $\tilde{\Omega}^{\prime}X^{\prime}$ and therefore does not modify the leading transverse growth rate $\sqrt{X^*}$.

\subsection{Finite-LSC fixed points}

Equations.~\eqref{eq:momentum_XY}--\eqref{eq:Y} also admit a pair of symmetry-related
finite-LSC fixed points,
\begin{equation}
    (\tilde{\Omega},X,Y)
    =
    (\pm\tilde{\Omega}_0,
    X_0,
    \pm Y_0),
    \qquad
    \tilde{\Omega}, X_0, Y_0>0.
\label{eq42}
\end{equation}
The positive branch corresponds to one LSC direction, while the negative branch corresponds to the LSC in the opposite direction.

For the positive branch, the fixed-point conditions are
\begin{equation}
    Y_0
    =a\tilde{\Omega}_0^{3/2},
    \qquad
    X_0
    =ab\tilde{\Omega}_0,
    \label{eq46}
\end{equation}
and
\begin{equation}
    \tilde{\Omega}_0^2
    +b^2\tilde{\Omega}_0
    -\frac{bc}{a}=0.
    \label{eq47}
\end{equation}
The physically relevant positive root is
\begin{equation}
    \tilde{\Omega}_0
    =
    \frac{
    \sqrt{b^4+4bc/a}-b^2
    }{2}.
\label{eq48}
\end{equation}
The thermal-dipole amplitude and orientation are therefore
\begin{align}
    \tilde{\xi}_0
    &=
    \sqrt{X_0^2+Y_0^2}
    =
    \sqrt{abc\tilde{\Omega}_0}, \label{eq51}\\
    \tan\phi_0
    &= \frac{Y_0}{X_0}
    =
    \frac{\sqrt{\tilde{\Omega}_0}}{b}. \label{eq52}
\end{align}
The negative-circulation state has the same $\tilde{\xi}_0$ and the opposite $\phi_0$.

For fixed $Pr$ and $N$, $a,b\propto Ra^{-1/4}$, whereas $bc/a$ is
independent of $Ra$.
Hence, as $Ra\rightarrow\infty$,
\begin{align}
     \tilde{\Omega}_0
    =
    \sqrt{\frac{bc}{a}}
    +O(Ra^{-1/2}).
    \label{eq53}
\end{align}
For $\Gamma=0.106$,
It follows that
\begin{align}
    \tilde{\Omega}_0
    &\simeq0.31Pr^{-1/3}, \label{eq54} \\
    \tilde{\xi}_0
    &\simeq
    5.93Ra^{-1/4}Pr^{-1/4}. \label{eq56}\\
    \tan\phi_0
    &\simeq \frac{0.093}{N} (RaPr)^{1/4}, \label{eq57}
\end{align}
These are the asymptotic fixed-point scalings given in Eq.~(12)--(14) of the main text.

\subsection{Eigenmodes about a finite-LSC state}

The two finite-LSC states are related by
$(\tilde{\Omega},X,Y)\rightarrow(-\tilde{\Omega},X,-Y)$ and therefore
have identical eigenvalues.  Evaluating Eq.~\eqref{eq:jacobian} at the
positive fixed point and using Eq.~\eqref{eq46} gives
\begin{equation}
    J_0=
    \begin{pmatrix}
        -\dfrac{3}{2}a\sqrt{\tilde{\Omega}_0}
        & 0
        & 1 \\[2mm]
        -\dfrac{1}{2}a\tilde{\Omega}_0^{3/2}
        & -b\sqrt{\tilde{\Omega}_0}
        & -\tilde{\Omega}_0 \\[2mm]
        \dfrac{1}{2}ab\tilde{\Omega}_0
        & \tilde{\Omega}_0
        & -b\sqrt{\tilde{\Omega}_0}
    \end{pmatrix}.
\label{eq59}
\end{equation}
The eigenvalues $\sigma$ are determined by
$\det(\sigma I-J_0)=0$, yielding
\begin{equation}
    \sigma^3+\mathcal{A}\sigma^2
    +\mathcal{B}\sigma+\mathcal{C}=0,
\label{eq60}
\end{equation}
where
\begin{align}
    \mathcal{A}
    &=
    \left(\frac{3}{2}a+2b\right)
    \sqrt{\tilde{\Omega}_0},
\label{eq61}\\
    \mathcal{B}
    &=
    \tilde{\Omega}_0^2+
    \left(\frac{5}{2}ab+b^2\right)
    \tilde{\Omega}_0,
\label{eq62}\\
    \mathcal{C}
    &=
    a\tilde{\Omega}_0^{3/2}
    \left(2\tilde{\Omega}_0+b^2\right).
\label{eq63}
\end{align}
Introducing the depressed-cubic coefficients
\begin{align}
    P
    &=
    \mathcal{B}-\frac{\mathcal{A}^2}{3}
    =
    \tilde{\Omega}_0^2
    -\frac{\tilde{\Omega}_0}{12}
    \left(9a^2-6ab+4b^2\right),
\label{eq64}\\
    Q
    &=
    \mathcal{C}
    -\frac{\mathcal{A}\mathcal{B}}{3}
    +\frac{2\mathcal{A}^3}{27}
    =
    \tilde{\Omega}_0^{5/2}
    \left(\frac{3}{2}a-\frac{2}{3}b\right)
    +
    \tilde{\Omega}_0^{3/2}
    \left(
    \frac{a^3}{4}
    -\frac{a^2b}{4}
    +\frac{ab^2}{6}
    -\frac{2b^3}{27}
    \right),
\label{eq65}
\end{align}
and the discriminant,
\begin{equation}
    S=
    \left(\frac{Q}{2}\right)^2
    +\left(\frac{P}{3}\right)^3.
\label{eq66}
\end{equation}
Over the experimental parameter range, $S>0$, yielding one real eigenvalue and one complex-conjugate pair:
\begin{align}
    \sigma_1
    &=
    -\frac{\mathcal{A}}{3}
    +
    \left(-\frac{Q}{2}+\sqrt{S}\right)^{1/3}
    +
    \left(-\frac{Q}{2}-\sqrt{S}\right)^{1/3},
\label{eq67}
\end{align}
and
\begin{align}
    \sigma_{2,3}
    &=
    -\frac{\mathcal{A}}{3}
    -\frac{1}{2}
    \left[
    \left(-\frac{Q}{2}+\sqrt{S}\right)^{1/3}
    +
    \left(-\frac{Q}{2}-\sqrt{S}\right)^{1/3}
    \right]
    \pm i\frac{\sqrt{3}}{2}
    \left[
    \left(-\frac{Q}{2}+\sqrt{S}\right)^{1/3}
    -
    \left(-\frac{Q}{2}-\sqrt{S}\right)^{1/3}
    \right],
\label{eq68}
\end{align}
where the cube roots are taken to be real.

For fixed $Pr$, both $a$ and $b$ scale as $Ra^{-1/4}$,
whereas $\tilde{\Omega}_0$ approaches an $Ra$-independent value.
Consequently, in the large-$Ra$ limit one has
\begin{align}
    P
    &=
    \tilde{\Omega}_0^2
    +\mathcal{O}(Ra^{-1/2}),
\label{eq69}\\
    Q
    &=
    \tilde{\Omega}_0^{5/2}
    \left(\frac{3}{2}a-\frac{2}{3}b\right)
    +\mathcal{O}(Ra^{-3/4}).
\label{eq70}
\end{align}
It follows that
\begin{align}
    \sqrt{S}=\sqrt{\left(\frac{Q}{2}\right)^2+\left(\frac{P}{3}\right)^3}
    \simeq \sqrt{\frac{\tilde{\Omega}_0^6}{27}+\mathcal{O}(Ra^{-1/2}})
    \simeq \frac{\tilde{\Omega}_0^3}{3\sqrt{3}}+\mathcal{O}(Ra^{-1/2}). \label{eq71}
\end{align}
To first order in $Ra^{-1/4}$,
\begin{align}
    & \left(-\frac{Q}{2}+\sqrt{S}\right)^{1/3}
    \simeq S^{1/6}-\frac{S^{-1/3}}{6}Q+\mathcal{O}(Q^2)
    \simeq
    \frac{\tilde{\Omega}_0}{\sqrt{3}}
    - \sqrt{\tilde{\Omega}_0}\left(\frac{3}{4}a-\frac{1}{3}b\right)+\mathcal{O}(Ra^{-1/2}), \label{eq72}\\
    & \left(-\frac{Q}{2}-\sqrt{S}\right)^{1/3}
    \simeq -S^{1/6}-\frac{S^{-1/3}}{6}Q+\mathcal{O}(Q^2)
    \simeq
    - \frac{\tilde{\Omega}_0}{\sqrt{3}}
    - \sqrt{\tilde{\Omega}_0}\left(\frac{3}{4}a-\frac{1}{3}b\right)+\mathcal{O}(Ra^{-1/2}). \label{eq73}
\end{align}
Substitution into Eqs.~\eqref{eq67} and \eqref{eq68} gives
\begin{align}
    \lim_{Ra\rightarrow\infty}
    \sigma_1
    =
    -2a\sqrt{\tilde{\Omega}_0}
    \simeq
    -38.1Ra^{-1/4}Pr^{1/12}, \label{eq74} \\
    \lim_{Ra\rightarrow\infty}
    \sigma_{2,3} = \sigma_r+i\sigma_i
    =
    \left(\frac{a}{4}-b\right)
    \sqrt{\tilde{\Omega}_0}
    \pm i\tilde{\Omega}_0.   \label{eq75}
\end{align}
These are the asymptotic eigenvalue scalings given in Eqs.~(16) and (17) of the main text.


\bibliography{references}